\documentclass[preprint]{aastex}

\begin{document}

\title{[NeIII]/[OII] as an Ionization Parameter Diagnostic in Star-Forming Galaxies}

\author{Emily M. Levesque\altaffilmark{1}}
\affil{CASA, Department of Astrophysical and Planetary Sciences, University of Colorado 389-UCB, Boulder, CO 80309}
\author{Mark L. A. Richardson}
\affil{School of Earth and Space Exploration, Arizona State University, Tempe, AZ 85287}
\altaffiltext{1}{Hubble Fellow; \texttt{Emily.Levesque@colorado.edu}}
\shortauthors{Levesque \& Richardson} 
\shorttitle{Ne3O2 in Star-Forming Galaxies}

\begin{abstract}
We present our parameterizations of the log([NeIII]$\lambda$3869/[OII]$\lambda$3727) (Ne3O2) and log([OIII]$\lambda$5007/[OII]$\lambda$3727) (O3O2) ratios as diagnostics of ionization parameter in star-forming galaxies. Our calibrations are based on the Starburst99/Mappings III photoionization models, which extend up to the extremely high values of ionization parameter found in high-redshift galaxies. While similar calibrations have been presented previously for O3O2, this is the first such calibration of Ne3O2. We illustrate the tight correlation between these two ratios for star-forming galaxies and discuss the underlying physics that dictates their very similar evolution. Based on this work, we propose the Ne3O2 ratio as a new and useful diagnostic of ionization parameter for star-forming galaxies. Given the Ne3O2 ratio's relative insensitivity to reddening, this ratio is particularly valuable for use with galaxies that have uncertain amounts of extinction. The short wavelengths of the Ne3O2 ratio can also be applied out to very high redshifts, extending studies of galaxies' ionization parameters out to $z\sim1.6$ with optical spectroscopy and $z\sim5.2$ with ground-based near-infrared spectra.
\end{abstract}

\maketitle

\section{Introduction} 
Quantifying the environmental properties of star-forming galaxies is crucial for our study of star formation and chemical evolution in the universe. Local galaxy samples can reveal the diversity of environments spanned by current star formation. As we extend these studies to progressively higher redshifts these same galaxies serve as laboratories for studying early star formation and its impact on metallicity evolution, the initial mass function, and reionization. Robust diagnostics utilizing the emission line spectra of star-forming galaxies can be used to unify our observations of these galaxies across a broad range of samples and redshifts. Calibrated using observational data, galaxy models, or a combination of the two, today's environmental diagnostics are our primary means of probing star-forming galaxies' stellar populations, star formation rates and histories, metallicities, and a number of other interstellar medium properties (e.g. Kennicutt 1998, Kewley et al.\ 2001, Kewley \& Dopita 2002, Pettini \& Pagel 2004, Tremonti et al.\ 2004, Dopita et al.\ 2006a). 

The emission line spectrum of a star-forming galaxy, produced by its HII regions, is driven by three interdependent parameters - the mean effective temperature of the hot massive star population (weighted by the number of ionizing photons), the metallicity ($Z$), and the ionization parameter (Dopita et al. 2006b). Ionization parameter, or $q$, is defined as the ratio of the mean ionizing photon flux to the mean atom density. Physically, $q$ can be described as the maximum velocity possible for an ionization front being driven by the local radiation field (and is sometimes expressed in its dimensionless form as $\mathcal{U} \equiv q/c$). This velocity $q$ is strongly dependent on metallicity (e.g. Evans \& Dopita 1985, Dopita \& Evans 1986). The ionizing photons in star-forming galaxies are produced by young hot massive stars, which have higher wind opacities and more efficient line blanketing at higher metallicities. As a result, the ionizing photons produced by the stars' photospheres are absorbed at a greater rate; some are more efficiently converted to mechanical energy in the stellar wind base region while a significant fraction are emitted at longer wavelengths (mainly near- and mid-infrared), leading to a net decrease in $q$ at higher $Z$ (Dopita et al. 2006a,b).

The correlation between metallicity and ionization parameter has proven challenging when trying to calibrate abundance diagnostics for star-forming galaxies and HII regions. Some metallicity diagnostics are only useful if the corresponding ionization parameter can be tightly constrained (e.g. McGaugh 1991, Kewley \& Dopita 2002). As a result, effective diagnostics of ionization parameter are crucial both for understanding the ionizing radiation field in a star-forming galaxy of HII region and for disentangling ionization parameter and metallicity effects in the use of abundance diagnostics. Currently, the most commonly-used ionization parameter diagnostic is log([OIII] $\lambda$5007/[OII]$\lambda$3727) (hereafter O3O2) (e.g. Alloin et al.\ 1978, Baldwin et al.\ 1981). However, the wavelength range spanned by the [OIII] and [OII] lines makes this ratio sensitive to extinction effects and decreases its efficacy in galaxies with a high or uncertain degree of extinction.

An intriguing alternative diagnostic for ionization parameter is the log([NeIII] $\lambda$3869/[OII]$\lambda$3727) ratio (hereafter Ne3O2). The similar short wavelengths of [NeIII] and [OII] make this ratio insensitive to reddening effects and usable as an ionization parameter diagnostic out to higher redshifts than O3O2 ($z\sim1.6$ as compared to $z\sim1$ for optical instruments and $z\sim5.2$ as compared to $z\sim3.8$ for ground-based near-IR; Nagao et al.\ 2006). Previous studies have presented this ratio as a metallicity diagnostic (Nagao et al.\ 2006, Shi et al.\ 2006). However, this is in fact an artifact of the $q$-$Z$ dependence given that neon closely tracks the oxygen abundance in star-forming galaxies and HII regions (Perez-Montero et al.\ 2007; for more discussion see Section 2). 

Here we present our parameterization of Ne3O2 as a useful alternative diagnostic of ionization parameter in star-forming galaxies. We consider the theoretical basis of this diagnostic's dependence on ionization parameter along with past work on the Ne3O2 diagnostic (Section 2). Combining an extensive grid of stellar population synthesis and photoionization models (Levesque et al.\ 2010 and Richardson et al.\ 2013) with observations of star-forming galaxies, we demonstrate the close correlation between Ne3O2 and O3O2 and present polynomial fits relating both ratios to ionization parameter. Finally, we discuss the current advantages and shortcomings of both diagnostics and consider the potential for future improvements that can be made with the next generation of star-forming galaxy models (Section 4).

\section{Neon in Emission Line Ratio Diagnostics}
Neon is produced during the late stages of massive stellar evolution. $^{20}$Ne is produced by carbon burning and is expected to closely track oxygen abundance; this correlation is confirmed by observations of extragalactic HII regions (Garnett 2002) and planetary nebulae (Henry 1989). Like oxygen, neon is abundant (solar $A$(Ne)$=8.11\pm0.04$; Cunha et al.\ 2006) and, being one of the dominant ionization species in HII regions, serves as one of the principal coolants along with oxygen (Burbidge et al.\ 1963, Gould 1963).

As a result of these properties, both singly-ionized and doubly-ionized neon have been previously utilized as tracers of the ionizing flux in star-forming galaxies. Ho \& Keto (2007) note that the high critical densities of [NeII] and [NeIII] make their fluxes insensitive to electron density ($n_e$) even in low-density HII regions and galaxies. The ratio of the [NeIII] and [NeII] fine-structure lines in the mid-IR have previous been utilized as probes of star formation rate, stellar population age, and the nature of the initial mass function and star formation history; O'Halloran et al. (2006) note that the mid-IR [NeIII]/[NeII] ratio is a very robust extinction- and abundance-insensitive indicator of the hardness of the radiation field surrounding massive young stars (see also Thornley et al.\ 2000, Perez-Montero et al.\ 2007). The ionization thresholds of Ne+ and Ne++ are 575\AA\ (ionization potential of 21.56 eV) and 303\AA\ (ionization potential of 40.96 eV) respectively, making the ratio of these two ionization states of neon a good direct tracer of the shape of the UV radiation field in mid-IR spectroscopy. This sensitivity has also been applied to the use of [NeIII] in diagnostics distinguishing star-forming galaxies and galaxies with AGN activity (e.g. Rola et al.\ 1997, Perez-Montero et al.\ 2007). 

With the similar abundance evolution of neon and oxygen, the optical-regime Ne3O2 ratio should be an even more powerful probe of the shape of the ionizing radiation field. The ionization thresholds of O+ and O++ are 911\AA\  (ionization potential of 13.62 eV) and 353\AA\ (ionization potential of 35.12 eV) respectively; the Ne3O2 ratio therefore spans a broader UV wavelength range than either O3O2 or the [NeIII]/[NeII] diagnostic, with a greater sensitivity at shorter wavelengths that accommodates more of the ionizing photons produced by young massive stars.

In addition, Ne++ is also abundant over a broader range of distances in the ionized nebula when compared to O++. Figure 1 shows the relative ionization fractions for neon and oxygen predicted by the Mappings III photoionization code, plotted as a function of relative distance from the inner surface of the nebula (for more discussion of the Mappings III code and nebular geometry see Section 3). As illustrated, the relative fraction of Ne++ is actual higher and extends to a greater distance than O++ (indeed, the relative ionization fraction of Ne++ and O+ become quite similar in the outer regions of the nebula). For oxygen, O+ begins as the dominant species in our low-density ($n_e \sim 100 cm^{-3}$) high-excitation model nebula but is eventually surpassed by neutral oxygen at greater distances from the inner nebular surface. With these two species dominating the ionization fraction, the O++ fraction is relatively low. However, in these same conditions Ne++ begins as the dominant ionic species and thus its ionization fraction persists to a greater distance in the nebula before eventually being overcome by Ne+; this has also been modeled and examined by Ho \& Keto (2007), whose results for the relative ionizing fractions are quite similar to ours (and see also O'Halloran et al. 2006). For a zero-age population of young massive stars, this should make the Ne3O2 ratio more sensitive to high ionization parameters (defined at the inner surface of the nebula in Mappings III) than O3O2. 

Nagao et al.\ (2006) were the first to present Ne3O2 as a potential metallicity diagnostic, arguing that the ratio's sensitivity to ionization parameter made it a good metallicity diagnostic as a result of the $q$-$Z$ dependence. Indeed, the Ne3O2 ratio does show a clear linear correlation with log(O/H)+12, as illustrated in both Nagao et al.\ (2006) and Shi et al.\ (2007). However, this correlation has a large residual scatter. Shi et al.\ (2007) present a linear fit for this metallicity calibration of log(O/H)+12$=-1.171(\pm0.008) \times {\rm Ne3O2} + 7.063(\pm0.727)$. They note that there is a substantial scatter in this relation and that its effective use as a metallicity diagnostic would require corrective calibrations such as direct measurements of the electron temperature and observations of additional emission features such as H$\beta$ and [OIII] $\lambda\lambda$4959,5007. Such requirements largely negate the benefits of Ne3O2 as a metallicity diagnostic, decreasing its efficacy as a short-wavelength (i.e. high redshift) diagnostic and requiring the use of other spectral features that serve as effective metallicity diagnostics in their own right (see, for example, Kewley \& Ellison 2008). Perez-Montero et al.\ (2007) further examine this large dispersion in the Ne3O2-metallicity relation, which equates to a standard deviation of 0.83 dex in the metallicity range. They conclude that these problems with the metallicity calibration of Ne3O2 stem from the ratio's primary dependence on ionization parameter. Perez-Montero et al.\ (2007) also note that Ne3O2 can be considered almost equivalent to O3O2 for diagnostic purposes, and that [NeIII] can be effectively substituted for [OIII] in many existing diagnostics. Combined, these previous studies present a compelling argument for calibrating Ne3O2 as an ionization parameter diagnostic.

\section{[NeIII]/[OII] as an Ionization Parameter Diagnostic}
For our ionization parameter diagnostic calibrations, we use the stellar population synthesis and photoionization models presented in Levesque et al.\ (2010) and Richardson et al.\ (2013). Both use the Starburst99 stellar population synthesis code (Leitherer et al.\ 1999, 2010) and the Mappings III photoionization code (Binette al.\ 1985, Sutherland \& Dopita 1993, Groves et al.\ 2004) to produce a grid of star-forming galaxy models.

Our Starburst99 models simulate a zero-age instantaneous burst of star formation with a fixed mass of $10^6M_{\odot}$ and a Salpeter initial mass function ($\alpha = 2.35$ for a mass range of 0.1M$_{\odot}$ - 100M$_{\odot}$; Salpeter 1955), allowing us to effectively model the ionization parameter in an active starburst galaxy. To model the atmospheres of the synthetic stellar population we adopt Starburst99's combination of the WMBASIC wind models of Pauldrach et al.\ (2001) and the CMFGEN Hillier \& Miller (1998) atmospheres for later ages dominated by Wolf-Rayet stars; both sets include rigorous non-LTE treatments of metal opacities and represent a substantial improvement over previous generations of diagnostic models (see Kewley et al.\ 2001, Levesque et al.\ 2010). The models span a range of five different metallicities ($Z = 0.001, 0.004, 0.008, 0.20$ and 0.040), produced by the adoption of Geneva ``high" mass loss stellar evolutionary tracks (Meynet et al.\ 1994). The final output from our Starburst99 models was a synthetic FUV spectrum produced using the isochrone synthesis method first introduced by Charlot \& Bruzual (1991).

The ionizing spectra generated by Starburst99 were then used as inputs for the Mappings III photoionization code. Using this ionizing spectrum, we used Mappings III to compute plane-parallel isobaric models with an adopted electron density of $n_e = 100$ cm$^{-3}$. Mappings III models the simulated photoionizated nebula as uniform nebular shells and computes the ionizing fraction of different species at each point through the nebula (for more discussion see Sutherland \& Dopita 1993, Groves et al.\ 2004). The ionization parameter $q$ is defined at the inner boundary of this nebula.

The Levesque et al.\ (2010) models span a range of ionization parameters from $1 \times 10^7$ cm s$^{-1} \le q \le 4 \times 10^8$ cm s$^{-1}$, chosen to agree with observed ionization parameters in local starburst galaxies (Rigby \& Rieke 2004). Richardson et al.\ (2013) extends this to higher ionization parameters, ranging from $6 \times 10^8$ cm s$^{-1}$ to the theoretical maximum of $q_{\rm max} = c$ (Groves et al.\ 2004). While it is true that local galaxies are generally not seen with $q > 3 \times 10^8$ cm s$^{-1}$, larger values of ionization parameter are expected in young extremely metal-poor starburst galaxies in the early universe undergoing their first cycles of star formation (e.g. Fosbury et al.\ 2003, Erb et al.\ 2010, Richard et al.\ 2011; for more discussion see Richardson et al.\ 2013). Models of high-redshift star-forming galaxies must therefore accommodate these larger values of ionization parameter. At these higher values of $q$, our models span only four values of metallicity, eliminating $z = 0.040$ to avoid modeling a non-physical parameter space.

The final output of Mappings III used in this work is a model emission-line spectrum that gives the intensities of all key emission lines relative to $H\beta = 1$. From this output we are able to determine the values of our emission line diagnostic ratios for each individually-modeled galaxy in our grid of metallicities (chosen in Starburst99) and ionization parameters (chosen in Mappings III).

Figure 2 shows the Ne3O2 and O3O2 diagnostic ratios predicted by our grid of star-forming galaxy models. The models illustrate the same close correlation that was described by Perez-Montero et al.\ (2007) and others. While there is some spread at the lower ionization parameters and highest metallicities ($\sim$0.4 dex) the tight relation between the two ratios demonstrates their effective equivalence as diagnostics. 

For comparison with our models, we also include several samples of star-forming galaxies in Figure 2. Our largest sample is drawn from Izotov et al.\ (2006), who examined the chemical composition of 309 metal-poor emission line galaxies from the Sloan Digital Sky Survey. We isolated this sample to objects that include [NeIII] and [OII] fluxes and satisfied the criteria for purely star-forming galaxies (that is, with no apparent AGN contribution) given in Kewley et al.\ (2001) and Kauffmann et al.\ (2003). This yielded a final sample of 107 star-forming galaxies. We also include a sample of six galaxies from Jaskot \& Oey (2013). These comprise the most strongly star-forming ``Green Pea" galaxies with the highest O3O2 ratios, presumed to indicate either an extremely high $q$ or an atypically low optical depth. Finally, we include galaxies from Xia et al.\ (2013), who obtained Magellan spectroscopy of three 0.2 $<z<$ 0.9 emission line galaxies selected from the HST/ACS Probing Evolution and Reionization Spectroscopically (PEARS) grism Survey.

It is clear from Figure 2 that the tight Ne3O2 vs. O3O2 correlation holds true for these observed star-forming galaxy samples as well as our models. However, there is a considerable offset (average $\sim$0.6 dex) between the models and the observations, corresponding to maximum offsets of $\lesssim -0.7$ dex for Ne3O2 or $\lesssim$ -0.9 dex for O3O2. Such an offset suggests that while the models do successfully reproduce the Ne3O2 vs. O3O2 correlation, there appears to be some underlying deficiency in their predicted emission line fluxes. Kewley et al.\ (2001) and Levesque et al.\ (2010) have previously discussed similar shortcomings with models produced by the Starburst99 and Mappings III codes, and conclude that they are the product of an insufficiently hard ionizing radiation spectrum. It is therefore unsurprising that a considerable offset would be apparent for the Ne3O2 and O3O2 diagnostic ratios, both of which span a broad range of ionizing wavelengths. While this does not diminish their efficacy as diagnostics of ionization parameter, the  offset shown in Figure 2 does effectively illustrate the need for additional improvements to the models used in developing such diagnostics, and we discuss this topic further in Section 4.

The direct relations between these diagnostic ratios and ionization parameter are illustrated in Figure 3 along with our best fits for the diagnostics. The polynomial fits were determined by using a polynomial regression of cubic order to fit the model grid points (for each metallicity set) of line diagnostic and ionization parameter. We quantified the goodness of fit using the coefficient of determination, $r^2$. The coefficients are defined such that: 
\begin{equation}
R = k_0+k_1x+k_2x^2+k_3x^3
\end{equation}

where R (ratio) corresponds to either Ne3O2 or O3O2, and $x=\log_{10}(q/\mbox{cm s}^{-1})$. Our fits accommodate the entire range of ionization parameter, but approach this with two separate fits for $q\le4\times10^8$ cm s$^{-1}$ and $q\ge4\times10^8$ cm s$^{-1}$ (with the exception of the z=0.040 metallicity, where the models only extend to $q=4\times10^8$ cm s$^{-1}$). These piecewise fits were done to accommodate for the apparent turnover in the diagnostic ratios at $q \sim 4\times10^8$ cm s${-1}$ (see Figure 3) and improve the goodness of the fits. Coefficients for these fits are given in Table 1. For comparison, we also plot the calibration fits presented in Kewley \& Dopita (2002) for O3O2 at comparable metallicities in Figure 3; based on this it is clear that our models have significantly reduced the spread in the O3O2 diagnostic as a function of metallicity.

Figure 3 illustrates that both the Ne3O2 and O3O2 ratios increase for higher values of $q$ but become less sensitive to $q$ at the highest values. Table 1 confirms that the $r^2$ values are comparable for both ratios. Both ratios show a dependence on abundance that becomes more apparent at the highest metallicities and ionization parameters. As originally noted in Kewley \& Dopita (2002), use of these ratios as ionization parameter diagnostics requires an initial guess of metallicity in order to effectively constrain $q$; however, Figure 3 also illustrates that for sufficiently high values (O3O2 $\gtrsim$ 1.2 and Ne3O2 $\gtrsim$ 0.6) the diagnostics ratios can be used to at least place limits on a maximum metallicity. Combined, Table 1 and Figure 3 demonstrate that Ne3O2 is just as robust a diagnostic of ionization parameter as O3O2 across a very broad range of values for $q$.

\section{Discussion}
We have presented our calibration of the Ne3O2 and O3O2 ratios as ionization parameter diagnostics based on the Starburst99/Mappings III photoionization models for star-forming galaxies from Levesque et al.\ (2010) and Richardson et al.\ (2013). The goodness of the fits to the model data for these two diagnostic calibrations demonstrate that Ne3O2 is an excellent alternative diagnostic to O3O2. This is further confirmed by the clear tight relation between the Ne3O2 and O3O2 diagnostics shown in the models as well as in observed samples of star-forming galaxies.

It is important to consider that, as illustrated by Figure 2, the calibrations presented here are based on models that cannot fully accommodate observations of the Ne3O2 and O3O2. The synthetic ionizing spectra are too weak and produce insufficient line fluxes, and it is unclear whether the disagreement in Figure 2 is primarily attributable to weak [NeIII] fluxes, weak [OIII] fluxes, or a combination of the two. This problem was previously discussed by Kewley et al.\ (2001) and Levesque et al.\ (2010), who both note that this same shortcoming is present in the models used to calibrate many of our current environmental diagnostics for star-forming galaxies. Improvements to stellar population synthesis and photoionization models should in turn lead to updated diagnostic calibrations that show better agreement with observed line fluxes in star-forming galaxies. For example, population synthesis models adopting stellar evolutionary tracks that include a detailed treatment of rotation effects (such as those presented in Ekstr\"{o}m et al.\ 2012 and Georgy et al.\ 2013) produce significantly harder ionizing radiation spectra (Levesque et al.\ 2012). This in turn should lead to stronger synthetic emission line fluxes, particularly for high-threshold species such as [NeIII] and [OIII], and improve the agreement between models and observations.

It is currently not possible to determine a correction between the Ne3O2 and O3O2 diagnostics due to complicating parameters such as metallicity, extinction, and the intrinsic spread between the observed Ne3O2 vs. O3O2 relation shown in Figure 2 ($\sim$0.4 dex). However, in general the Ne3O2 diagnostic calibration yields slightly lower ionization parameters than the O3O2 calibration, with maximum disagreements of a factor of 2 to 5. The disagreement between the two diagnostics is worse at higher metallicities, implying that we should find better agreement between these diagnostics for high-redshift (lower-metallicity) galaxies. It is also likely that agreement between the two diagnostics will also improve as improved photoionization models with more realistic emission line fluxes become available.

With two diagnostic calibrations of such similar quantity, it is worth discussing specific scenarios or data samples in which Ne3O2 or O3O2 would be recommended as the preferred ionization parameter diagnostic. For observations of galaxies with weak emission line fluxes or a poor S/N spectrum, O3O2 is likely to remain the preferred diagnostic given the typically much stronger fluxes of the [OIII] $\lambda$5007 emission line as compared to the [NeIII] $\lambda$3869 emission line. However, we can see from Figure 3 that at higher ionization parameters ($q \gtrsim 10^8$ cm s$^{-1}$) the [NeIII] $\lambda$3869 and [OII] $\lambda$3727 fluxes become comparable, which should make Ne3O2 just as viable a diagnostic as O3O2 for such galaxies even with low-S/N spectra. In addition, for galaxies with uncertain amounts of extinction, the Ne3O2 ratio is also a preferable choice given its lower sensitivity to reddening effects. Finally, the Ne3O2 ratio can be applied to observations extending to higher redshifts than O3O2; optical spectroscopy of $1 < z < 1.6$ galaxies and ground-based NIR spectroscopy of galaxies with $3.8 < z < 5.2$ will only offer coverage of the Ne3O2 ratio. \\

We thank Maria Pe\~{n}a-Guerrero and Lifang Xia for valuable discussions and correspondence regarding this work, as well as the anonymous referee for constructive comments that improved the content of this manuscript. EML is supported by NASA through Hubble Fellowship grant  number HST-HF-51324.01-A from the Space Telescope Science Institute, which is operated by the Association of Universities for Research in Astronomy, Incorporated, under NASA contract NAS5-26555. MLAR was supported by the National Science and Engineering Research Council of Canada.

\begin{figure}
\epsscale{1}
\plotone{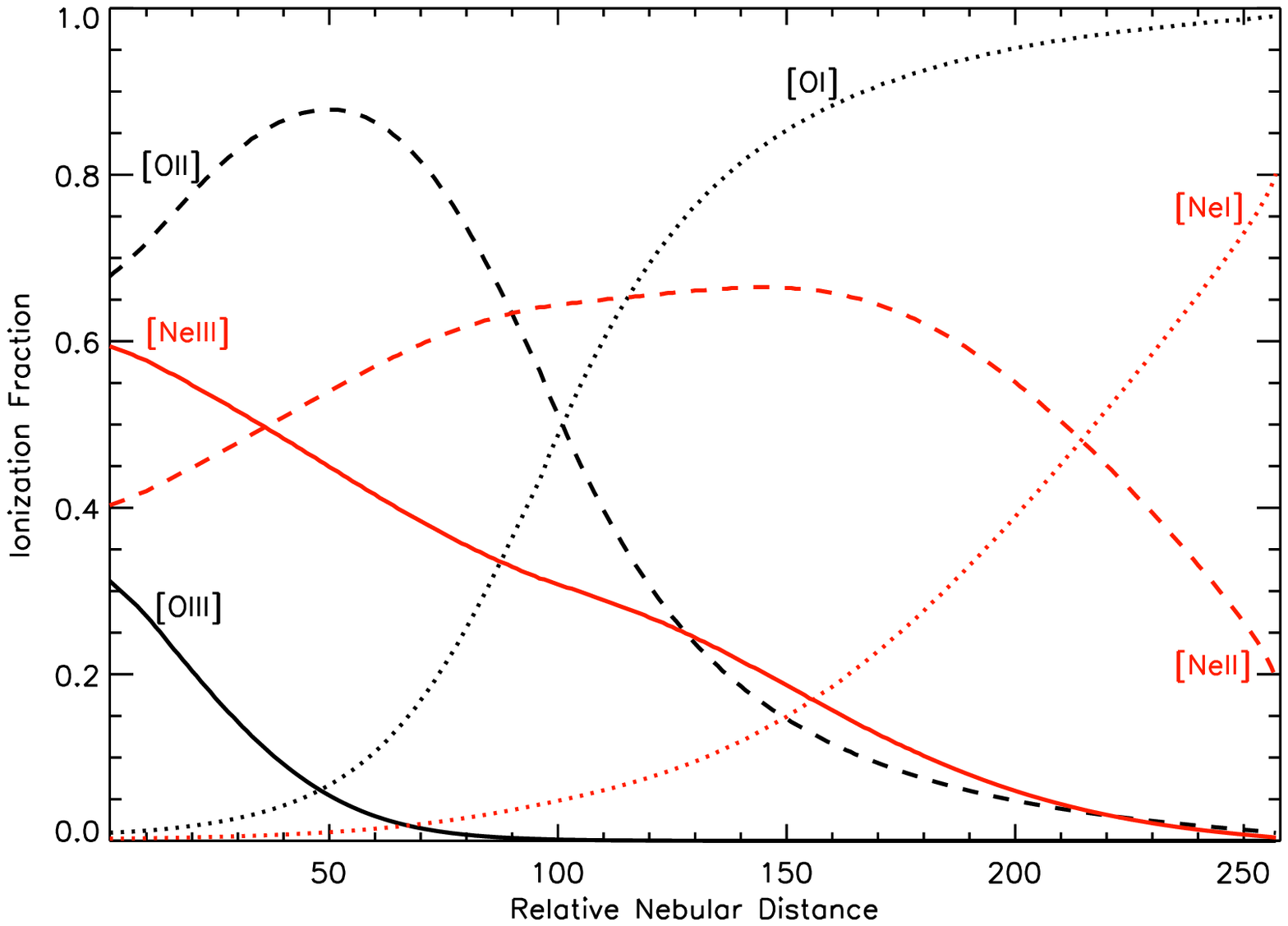}
\caption{Relative ionization fractions for neutral (dotted), singly-ionized (dashed), and doubly-ionized (solid) oxygen (black) and neon (red) produced by our Mappings III models, plotted as a function of relative distance from the inner surface of the nebula (defined here as a count progression through the series of uniform nebular shells defined by Mappings III that make up the model nebula; see Groves et al. 2004). For this plot a metallicity of $Z = 0.001$, an ionization parameter of $q = 1 \times 10^{7}$ cm s$^{-1}$, a zero-age instantaneous burst star formation history, and an electron density of $n_e = 100$ cm$^{-3}$ is assumed. }
\end{figure}

\begin{figure}
\plotone{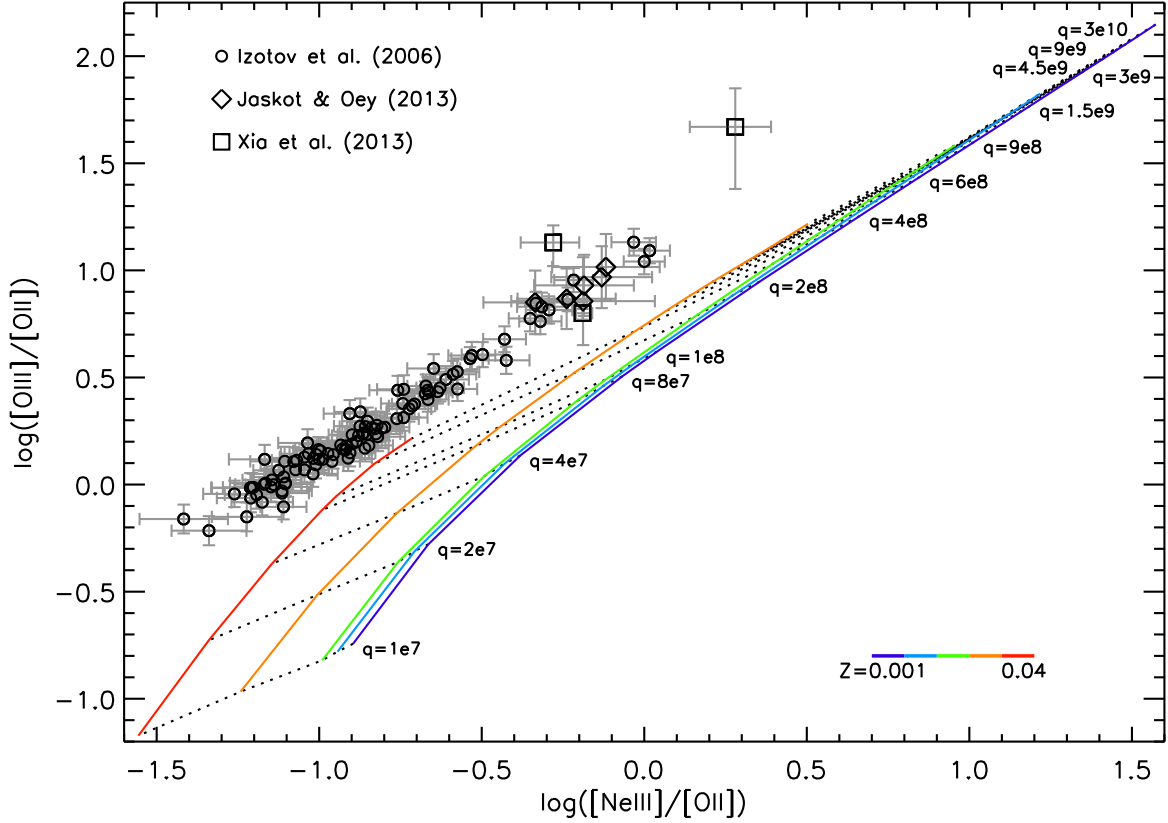} 
\caption{[NeIII]/[OII] vs. [OIII]/[OII] from our grid of photoionization models, assuming a zero-age instantaneous burst star formation history and an electron density of $n_e = 100$ cm$^{-3}$. The models are plotted with lines of constant metallicity (colored solid lines as indicated in the legend) and ionization parameter (labeled black dotted lines). The model grid is compared to a sample of 107 star-forming galaxies from Izotov et al.\ (2006; circles), 6 from Jaskot \& Oey (2013; diamonds), and 3 from Xia et al.\ (2013; squares). Errors are shown in gray.}
\end{figure}

\begin{figure}
\epsscale{0.49}
\plotone{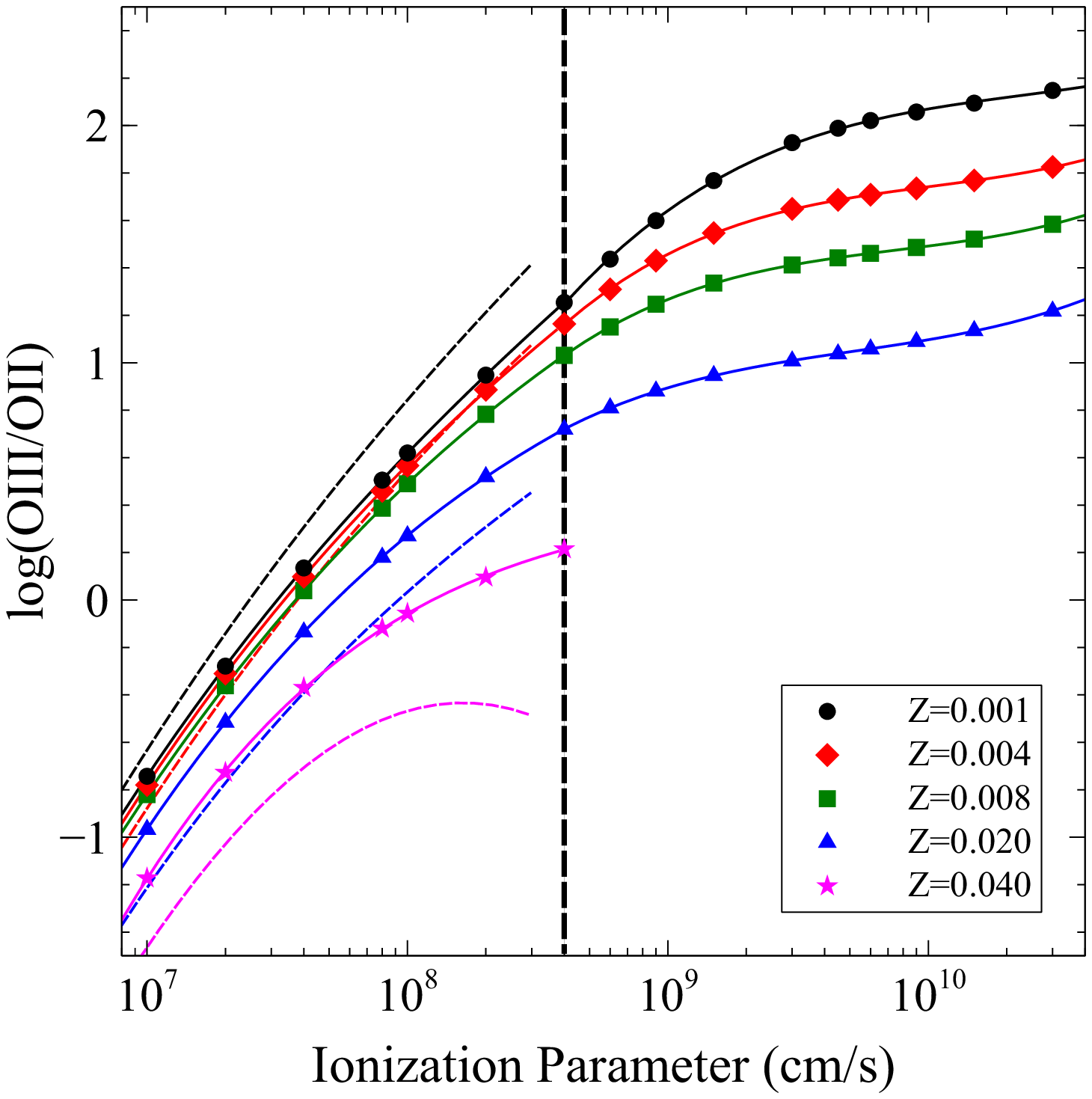}
\plotone{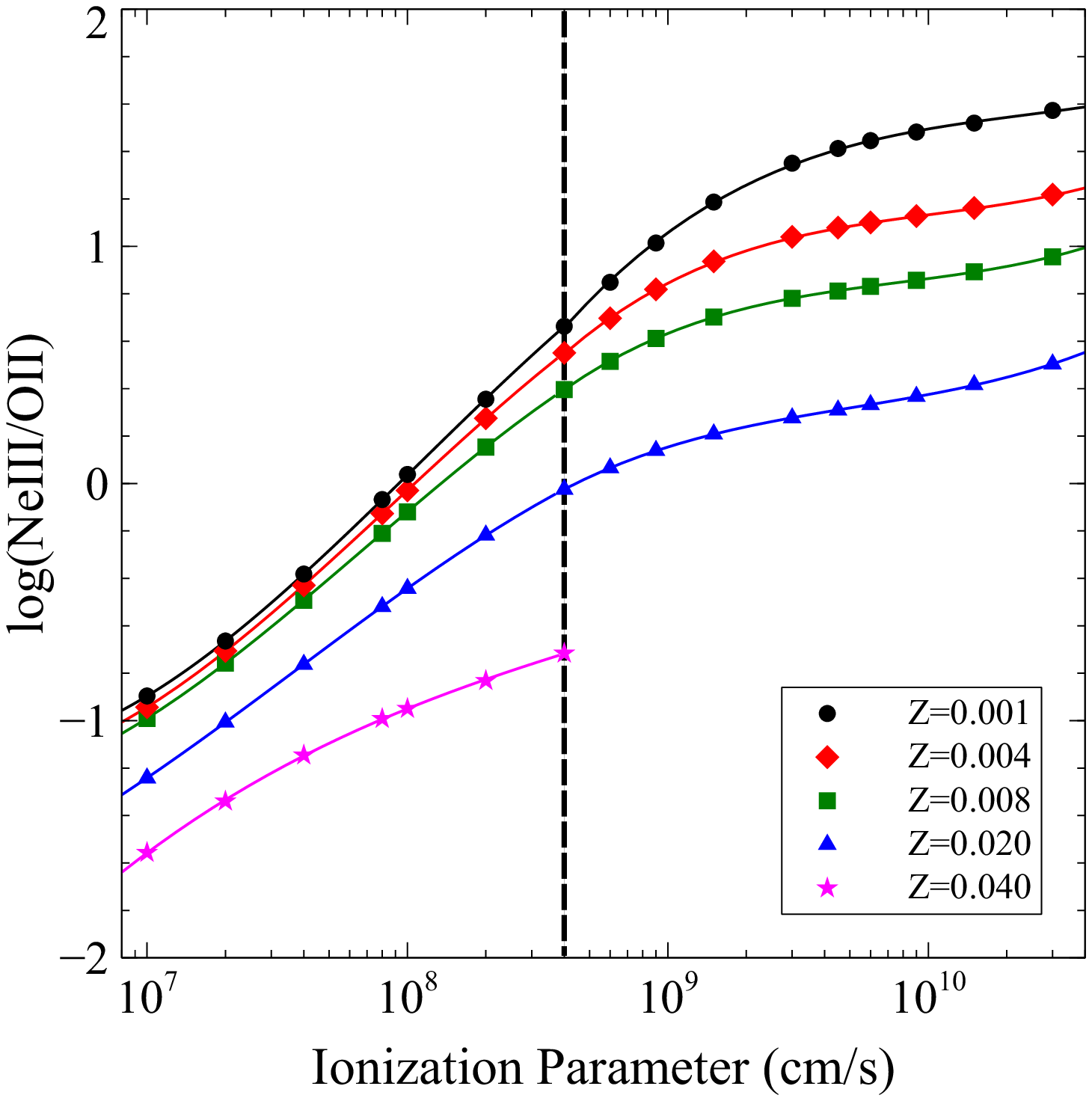}
\caption{The O3O2 diagnostic (left) and Ne3O2 diagnostic (right) vs. ionization parameter. Curves for each model metallicity are fit to the data using a cubic polynomial regression (see Table 1 for coefficients) and indicated by distinct colors and symbol shapes. We fit the $q\le 4 \times 10^8$ cm s$^{-1}$ points with one cubic fit and the $q\ge 4 \times 10^8$ cm s$^{-1}$ points with a second cubic fit. Dashed lines in the O3O2 diagnostic plot illustrate polynomial fits to this diagnostic from Kewley \& Dopita (2002).}
\end{figure}

\begin{deluxetable}{llrrrrr}
\tabletypesize{\scriptsize}
\tablecaption{Coefficients for Ionization Parameter Diagnostic Fits
\label{qcoefftable1}}
\tablehead{{Diagnostic (R)}
&
& {Z=0.001}
& {0.004}
& {0.008}
& {0.020}
& {0.040}\\}
\startdata
$\log\left(\frac{[{\rm NeIII}]}{[{\rm OII}]}\right)$ & \multicolumn{6}{l}{
($1.0 \times 10^7$ cm s$^{-1} \le q \le 4.0 \times 10^{8}$ cm s$^{-1}$)}\\
  & $r^2$ & 0.999994 & 0.999995 & 0.999999 & 0.999997 & 0.999843 \\
  & $k_{0}$ &  56.3416 &  53.5278 & 48.7182 & 24.7917 &  -33.9635  \\
  &  $k_{1}$ &  -23.1202  & -22.2764 &  -20.6263 & -11.7739 & 10.5159 \\
  &  $k_{2}$ &  3.00640 & 2.93137 & 2.75080 & 1.66339 & -1.13693 \\
  &  $k_{3}$ & -0.124519 & -0.122954 & -0.116949 & -0.0732419 & 0.0422855 \\ \hline
$\log\left(\frac{[{\rm NeIII}]}{[{\rm OII}]}\right)$ & \multicolumn{6}{l}{
($4.0 \times 10^8$ cm s$^{-1} \le q \le 3.0 \times 10^{10}$ cm s$^{-1}$)}\\
  & $r^2$ & 0.999686 & 0.999933 & 0.999973 & 0.999962 & \nodata \\
  & $k_{0}$ & -129.649 &  -135.261 & -122.700 & -97.2988 &  \nodata  \\
  &  $k_{1}$ &  37.6573  & 40.5373 &  37.1900 & 29.7713 & \nodata \\
  &  $k_{2}$ &  -3.61621 & -4.03016 & -3.74568 & -3.04451 & \nodata \\
  &  $k_{3}$ & 0.116192 & 0.134037 & 0.126232 & 0.104412 & \nodata \\ \hline \hline
$\log\left(\frac{[{\rm OIII}]}{[{\rm OII}]}\right)$ & \multicolumn{6}{l}{
($1.0 \times 10^7$ cm s$^{-1} \le q \le 4.0 \times 10^{8}$ cm s$^{-1}$)}\\
  & $r^2$ & 0.999999 & 0.999987 & 0.999997 & 0.999996 & 0.999900 \\
  & $k_{0}$ &  -45.5266 &  -46.4132 & -40.2204 & -50.3083 &  -87.3717  \\
  &  $k_{1}$ &  13.6587  & 13.7967 &  11.2433 & 14.8567 & 28.4997 \\
  &  $k_{2}$ &  -1.39412 & -1.38377 & -1.03169 & -1.45589 & -3.11096 \\
  &  $k_{3}$ & 0.0509756 & 0.0491601 & 0.0327995 & 0.0486398 & 0.114098 \\ \hline
$\log\left(\frac{[{\rm OIII}]}{[{\rm OII}]}\right)$ & \multicolumn{6}{l}{
($4.0 \times 10^8$ cm s$^{-1} \le q \le 3.0 \times 10^{10}$ cm s$^{-1}$)}\\
  & $r^2$ & 0.999706 & 0.999941 & 0.999968 & 0.999913 & \nodata \\
  & $k_{0}$ &  -128.151 &  -135.392 & -124.662 & -100.309 &  \nodata  \\
  &  $k_{1}$ &  37.4523  & 40.7980 &  38.0220 & 30.9800 & \nodata \\
  &  $k_{2}$ &  -3.60270 & -4.05994 & -3.83393 & -3.17249 & \nodata \\
  &  $k_{3}$ & 0.115968 & 0.135148 & 0.129327 & 0.108855 & \nodata \\ \hline
\enddata
\end{deluxetable}




\end{document}